\documentclass{appolb}
\usepackage{graphicx}
\usepackage[left]{lineno} 

\begin{document}
\title{Low-x QCD at the LHC with the ALICE detector%
\thanks{Presented at Excited QCD 2009, winter meeting on QCD, 8-14 February 09, Zakopane, Poland}%
}
\author{Magdalena Malek\thanks{Magdalena.Malek@cern.ch} for the ALICE Collaboration
\address{Institut de Physique Nucl\'{e}aire d'Orsay (IPNO) - France\\
CNRS: UMR8608 - IN2P3 - Universit\'{e} Paris Sud - Paris XI}
}
\maketitle
\begin{abstract}
We give a brief review of the physics of gluon saturation and non-linear QCD evolution at small
values of the Bjorken-$x$ variable. We discuss the ALICE capability for low-$x$ studies at the
LHC. In particular, we concentrate on the heavy quark production in the CGC framework and its
observation with the ALICE Muon Spectrometer.
\end{abstract}

\section{Heavy quark production cross section} The heavy quark production cross
section in $H_{1}$-$H_{2}$ collisions can be illustrated by the formula:
\begin{equation}
\sigma^{H_{1}, H_{2}} = \sum_{i,j}\int
dx_{1}dx_{2}f_{i}(x_{1})^{H_{1}}f_{j}(x_{2})^{H_{2}}\sigma_{ij}(x_{1},x_{2},s) \label{cs}
\end{equation}
where $\sigma_{ij}(x_{1},x_{2},s)$ is the perturbative partonic cross section for two
hard-scatterings to produce a pair of heavy quarks which depends in particular on the available
energy. This parton level cross section is calculable as a power series in the strong coupling
$\alpha_{s}$ which depends on the renormalisation scale. The nucleon Parton Distribution Functions
(PDF) $f_{i/j}(x_{1}/x_{2})^{H_{1}/H_{2}}$ gives the probability to find the parton $i$/$j$ in a
nucleon of type $H_{1}$/$H_{2}$ as a function of fraction the $x_{1}$/$x_{2}$ of the nucleon's
momentum carried by the parton. The PDFs are non-perturbative objects and thus cannot be
calculated from first principles. They are extracted from the structure functions of the nucleon
in deep inelastic scattering (DIS) experiments. The comprehension of PDFs is mandatory to make any
predictions about heavy flavor production.
\section{Parton structure and evolution equations}
\subsection{Parton distribution in linear regime}
The DGLAP
(\textbf{D}okshitzer-\textbf{G}ribov-\textbf{L}ipatov-\textbf{A}ltarelli-\textbf{P}arisi)~\cite{DGLAP}
evolution equation applicable in the Bjorken limit ($Q^{2}\rightarrow\infty$ and fixed $x$) takes
into account the $Q^{2}$ dependence of the PDFs. It allows a resummation of leading powers of
[$\alpha_{s}\ln(Q^{2})]^{n}$ generated by a parton cascade in a region of the phase space where
the transverse momenta of gluons are strongly ordered:
$Q^{2}~\gg~k_{nt}^{2}~\gg~k_{(n-1)t}^{2}~\gg~\ldots~\gg~k_{1t}^{2}$. In this approach the
transverse momenta of the initial partons are neglected.\\
At high energy (or decreasing $x$), the probability of gluon emission increases as
$\alpha_{s}\ln(1/x)$. In this regime, the transverse momenta of the partons are of the same order
of magnitude than the longitudinal ones and thus have to be taken into account in the reaction
dynamics description. When the gluon density $\mathcal{N}$ is not too high, the modification of
the gluon distribution with increasing energy can be described by the BFKL
(\textbf{B}alitsky-\textbf{F}adin-\textbf{K}uraev-\textbf{L}ipatov) evolution
equation~\cite{BFKL}. This equation can be used in the Regge limit ($s\rightarrow\infty$ and fixed
$Q^{2}$) of QCD. The PDFs are described in terms of $k_{t}-unintegrated$ PDFs $\phi(x,k_{t})$
related to the usual gluon distribution $xg(x,Q^{2})$ by
$xg(x,Q^{2})~\sim~\int_{0}^{Q^{2}}d^{2}k_{t}\phi(x,k_{t})$. This is essentially a linear equation
and its solution at asymptotically large energies shows an exponential growth of the number of
gluons. This unlimited growth of the gluon density leads ultimately to a violation of the
unitarity bound for the cross section. The BFKL equation must be modified so as not to violate the
unitarity.
\subsection{Non-linear evolution and parton saturation}
At some small enough value of $x$, the gluon density is so large that the non-linear gluon
recombination becomes important. The gluon density per unit area of the nucleus with atomic number
A is $\rho_{A}\sim xg_{A}(x,Q^{2})/A^{2/3}$ and the gluon recombination cross section is
$\sigma_{gg\rightarrow g}\sim\alpha_{s}/Q^{2}$. The border line between the linear and non-linear
regimes is given by the saturation scale $Q_{s}$ satisfying $\rho_{A}\sigma_{gg\rightarrow~g}=1$.
The gluon recombination becomes important when $\rho_{A}\sigma_{gg\rightarrow~g}>1$ thus
$Q^{2}\leq Q^{2}_{s}$ with $Q_{s}^{2}\sim\alpha_{s}xg_{A}(x,Q^{2}_{s})/A^{2/3}$. Partons with
$Q^{2}>Q^{2}_{s}$ are not affected by saturation, evolution goes towards the dilute system, and
can be described by DGLAP or BFKL equations. On the other hand, partons with $Q^{2}<Q^{2}_{s}$ are
in the saturation regime and their evolution is described by the non-linear JIMWLK
(\textbf{J}alilian
Marian-\textbf{I}ancu-\textbf{M}cLerran-\textbf{W}eigert-\textbf{L}eonidov-\textbf{K}ovner)
equation~\cite{CGCreviews}. This equation can be schematized by:
\begin{equation}
\frac{\partial \mathcal{N}(k_{t},y)}{\partial y} \simeq c_{1}\mathcal{N}(k_{t},y) -
\alpha_{s}c_{2}\mathcal{N}^{2}(k_{t},y) \label{eJIMWLK}
\end{equation}
with rapidity $y\equiv\log(1/x)$ and constant parameters $c_{1,2}$ of the order of one. One
remarks that the gluon recombination (or quadratic term) becomes important when $\mathcal{N}$
reaches 1/$\alpha_{s}$. The gluon production stops growing, this is the so called gluon
saturation. The saturation phenomena can be described by an effective theory: the \textbf{C}olor
\textbf{G}lass \textbf{C}ondensate (CGC). The CGC is a new form of nuclear matter which controls
hadronic interactions at asymptotically large energies. The origin of the name is the following.
First of all, the CGC is made of small-$x$ gluons which carry the \textbf{color} charge. Next,
these small-$x$ gluons are created by partons with larger $x$ which act as a frozen color source,
\textbf{glass}, for the emission of small-$x$ gluons. Lastly, \textbf{condensate} because the soft
degrees of freedom are as densely packed as they can.
\section{Low-x QCD studies at LHC with ALICE}
The LHC (\textbf{L}arge \textbf{H}adron \textbf{C}ollider) at CERN will provide, at nominal
operating conditions, p+p, Pb+p and Pb+Pb collisions at $\sqrt{s_{NN}}$~=~14, 8.8 and 5.5 TeV
respectively with luminosities $\mathcal{L}\sim~$10$^{34}$, 10$^{29}$ and 5$\times$10$^{26}$
cm$^{-2}$s$^{-1}$. These conditions allow to access unprecedentedly low values of the Bjorken-$x$
variable. The copious production of heavy quarks (charm and beauty) can be used as a probe of
low-$x$ QCD phenomena. At LHC energies heavy quarks are mainly produced through gluon-gluon fusion
processes thus their production cross sections
are significantly affected by parton dynamics in the small-$x$ regime.\\
ALICE (\textbf{A} \textbf{L}arge \textbf{I}on \textbf{C}ollider \textbf{E}xperiment)~\cite{ALICE}
is the experiment dedicated to study the properties of matter created in Pb+Pb collisions at the
LHC. The detector contains a wide array of sub-detectors for measuring hadrons, leptons and
photons. Table~\ref{8t1} shows the $x$ regimes for the production of $c\overline{c}$ and
$b\overline{b}$ pairs at SPS, RHIC and LHC (for ALICE) energies. As we can see in Tab.~\ref{8t1}
the increase of the collision energy involves the decrease of the value of the Bjorken-$x$.
\begin{table}[!h]
   \begin{center}
      \begin{tabular}{|c|c|c|c|c|}
    \hline
       Machine                                   &     SPS    &    RHIC      &  LHC & LHC \\
       System & Pb+Pb&Au+Au&Pb+Pb&p+p\\
       Energy~(GeV)                             &     17    &    200      &  5500 &  14000\\
    \hline
    \hline
\hline
    $c\overline{c}$    &     x $\simeq$ 10$^{-1}$  &  x $\simeq$ 10$^{-2}$ & x $\simeq$ 4 $\times$ 10$^{-4}$ &  x $\simeq$ 2 $\times$ 10$^{-4}$   \\
    $b\overline{b}$         &    -    &-         &      x $\simeq$ 2 $\times$ 10$^{-3}$    &x $\simeq$ 6 $\times$ 10$^{-4}$  \\
    \hline
    \end{tabular}
  \end{center}
   \caption{Bjorken-$x$ values for charm and beauty production at central rapidity and $p_{t}$~=~0 at SPS, RHIC
   and LHC energies. Table extracted from~\cite{ALICE}.}
  \label{8t1}
\end{table}

\subsection{Simulations: setup}
The simulation of the heavy quark production within the CGC framework is performed with the
AliRoot~\cite{aliroot} simulation code. The theoretical ingredients (PDF and partonic cross
section) of Eq.~\ref{cs} for each approach are discussed in next sections. The results are
compared to the reference simulations for the ALICE experiment (PYTHIA) that are based on the
\textbf{M}angano-\textbf{N}ason-\textbf{R}idolfi (MNR) calculation~\cite{MNR}. The comparison
between p+p and Pb+p collisions allows to draw preliminary conclusions about the CGC observation
with the ALICE muon spectrometer~\cite{muon} covering the forward pseudo-rapidity region
(-4$\leq\eta\leq$-2,5).
\subsubsection*{Collinear factorization: PYTHIA tuned to MNR}

\begin{itemize}
\item p+p collisions are simulated with PYTHIA tuned to reproduce heavy quark production cross
sections obtained with MNR calculation. The CTEQ~\cite{PumplBHST1} parametrization of the proton
PDF is used.
\item Pb+p collisions are simulated from p+p ones using Glauber geometric model~\cite{glau} to
calculate the number of binary collisions. Shadowing effects in the gluon distribution are added
by hand, using a nuclear modification factor. This factor, not well known at LHC energies, can be
calculated using different parametrization methods. In this study we used the EKS98
parametrization~\cite{EskolKS2}. Within this model, the transverse momentum $k_{t}$ of partons is
given by means of a Gaussian distribution with a zero mean value and a width ($\sigma$~=~$\mid
k_{t} \mid$) that depends on the considered flavor and the colliding system. For p+p and Pb+p
collisions, this parameter is respectively fixed to 1 and 1.16 GeV/c for charm, and 1 and 1.6
GeV/c for beauty.
\end{itemize}

\subsubsection*{CGC approach}
In this approach the proton PDF is given by the non-saturated CTEQ parametrization. The lead
structure is described by the PDF taking into account the saturation effects.
More precisely, it contains the multiple scattering corrections
in the partonic cross section and the non-linear terms of the JIMWLK equation. The initial
condition for the JIMWLK equation is calculated for $x$~=~10$^{-2}$ using the McLerran-Venugoplan
model~\cite{McLerV1}.
\subsection{Results on heavy quarks level}
We calculate the charm and beauty production cross section and their ratios for both approaches.
The ratio $R_{Pbp} = \frac{dN_{Pbp}/dydp_{t}}{N_{coll}dN_{pp}/dydp_{t}}$ defines the nuclear
modification factor that can be calculated from experimental data. The results are given in
Tab.~\ref{tHQRatios}. One remarks that the values of shadowing factors for beauty are very close
in both models. For charm, these factors are very different. It shows that the gluon recombination
effects are more important in the CGC approach that in the MNR one. In fact, the $c\overline{c}$
pairs are produced, on average, by fusion of gluons at smaller $x$ than the ones producing the
$b\overline{b}$ pairs ($x=\frac{M_{Q\overline{Q}}}{\sqrt{s_{NN}}} \exp(\pm y_{Q\overline{Q}})$).
Since small-$x$ gluons are more sensitive to recombination effects, these are more
visible in the charm case.\\
Table~\ref{tHQRatios} gives also the cross section ratios for beauty over charm production. One
sees that the MNR approach  always predicts a ratio smaller than 5\%, whereas the CGC approach
gives a ratio larger than 10\%.\\

\begin{figure}[ht]
\includegraphics*[width=6.5cm]{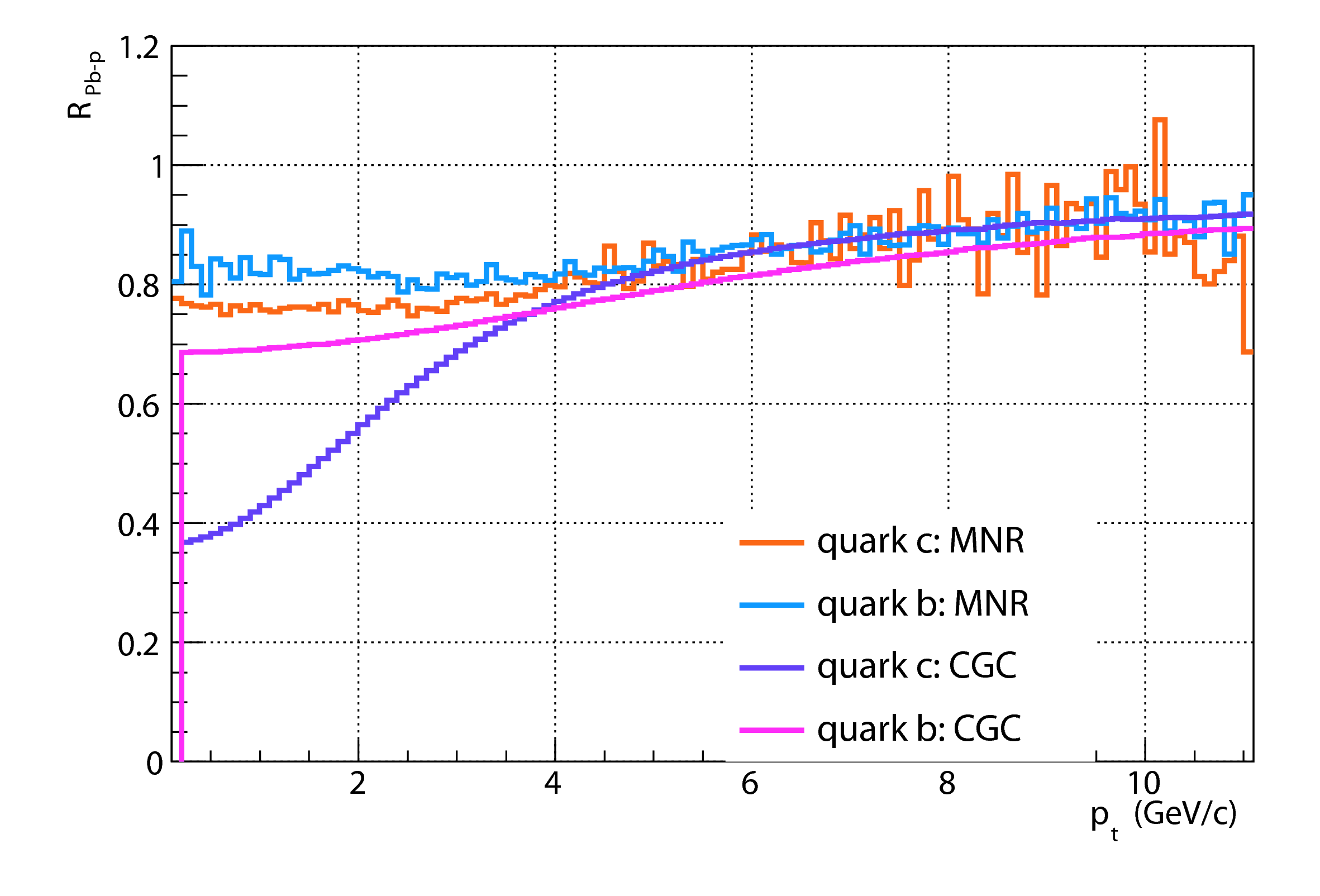}
\includegraphics*[width=6.5cm]{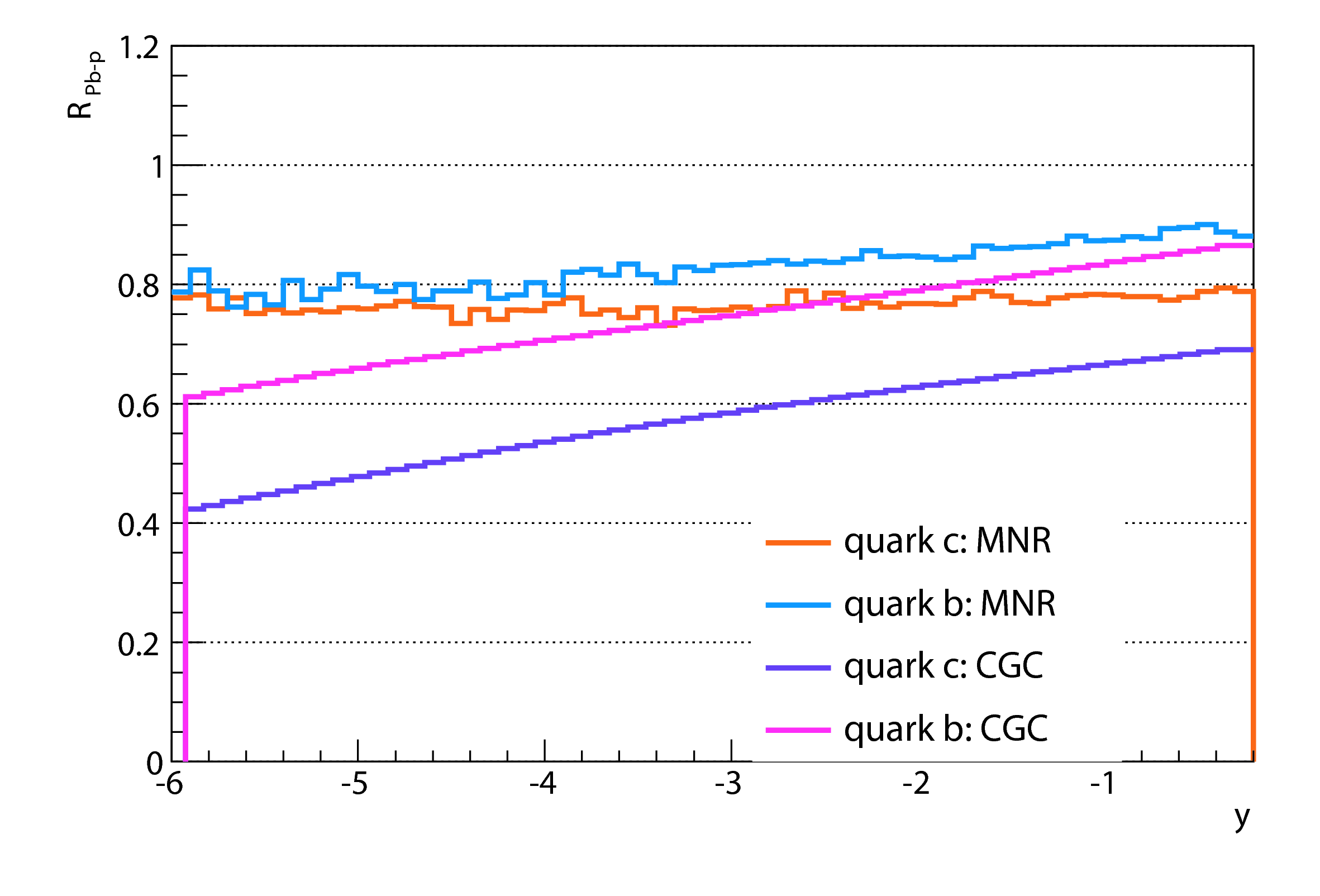}
\caption{$R_{Pbp}$ ratios as a function of $p_{t}$ (left part) and rapidity (right part).}
\label{Fig1}
\end{figure}

\begin{table}[!t]
\begin{center}
\begin{tabular}{|c|c|c|c|c|}
\hline
 approach& $dN_{Pbp}^{c\overline{c}}/dN_{pp}^{c\overline{c}}$ & $dN_{Pbp}^{b\overline{b}}/dN_{pp}^{b\overline{b}}$ &$dN_{pp}^{b\overline{b}}/dN_{pp}^{c\overline{c}}$&$dN_{Pbp}^{b\overline{b}}/dN_{Pbp}^{c\overline{c}}$\\
\hline
\hline \hline
MNR with EKS98 & 0.77&0.85&0.03&0.04\\
CGC & 0.60&0.80&0.10&0.14\\
\hline
\end{tabular}
\end{center}
\caption{Production ratios of charm and beauty in Pb+p collisions at $\sqrt{s_{NN}}$~=~8.8~TeV
normalized to the number of binary collisions $N_{coll}$ and cross section ratios between beauty
and charm for rapidity $y~<~0$ in MNR and CGC approaches.} \label{tHQRatios}
\end{table}
The evolution of the nuclear modification factor $R_{Pbp}$ as a function of transverse momentum
and rapidity is presented in Fig.~\ref{Fig1}.
One remarks that below $\sim$4~GeV/c, in the CGC approach, the depletion for charm is much more
important than for beauty. It can be understood by recombination and multiple scattering of
small-$x$ gluons which are responsible for charm production. This effect is less pronounced in the
MNR~+~EKS98 approach. Shadowing effects are also visible from the rapidity dependence of
$R_{Pbp}$. Both approaches show the decrease of this ratio at high rapidities but it is clearly
more noticeable, for both flavors, in the CGC approach.
\subsection{Results on hadron and muon level}
Next, the quark fragmentation into heavy hadrons was performed. The decay of produced hadrons
through their semi-muonic channel was applied. Finally, the transport of produced muons through
the muon spectrometer was completed~\cite{transport}. It is important to highlight that a full
AliRoot simulation was used to transport particles and to reconstruct their tracks; thus a
realistic signal generation from the full detector response and a complete clustering and tracking
procedure were
applied.\\
\indent Reconstructed single muon spectra are studied in this part. The shift of the center of
mass frame with respect to the laboratory one, by $\delta y = 0.47$ for Pb+p collisions, is taken
into account. This kinematical shift constrains the rapidity window where the $R_{Pbp}$ can be
measured; the ratio will correspond to muons with $y\in$~[-2.97,-4] for
p+p collisions and muons with $y \in$~[-2.5,-3.53] for Pb+p collisions.\\

\begin{figure}[ht]
\includegraphics*[scale=0.09]{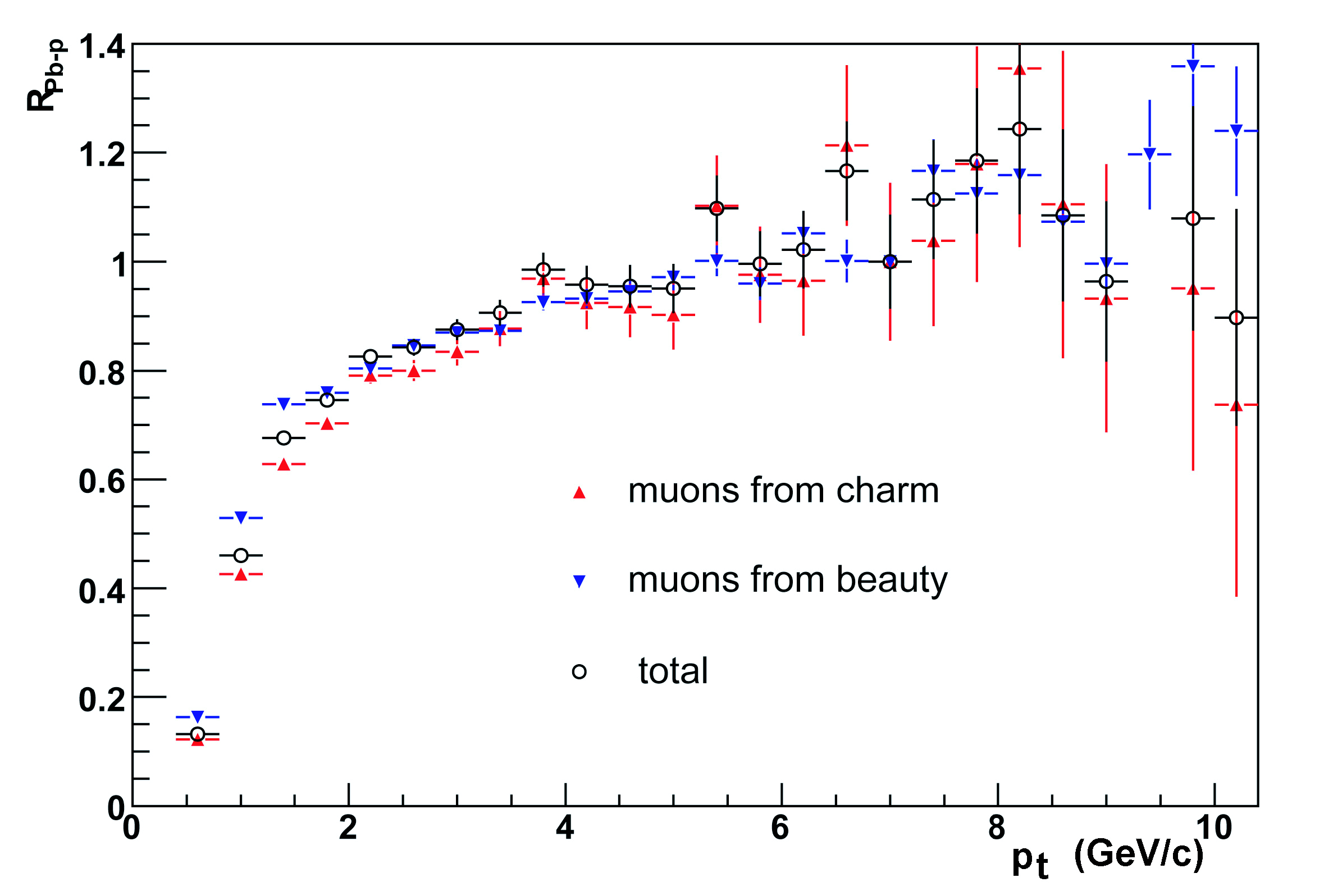}
\includegraphics*[scale=0.092, angle=90]{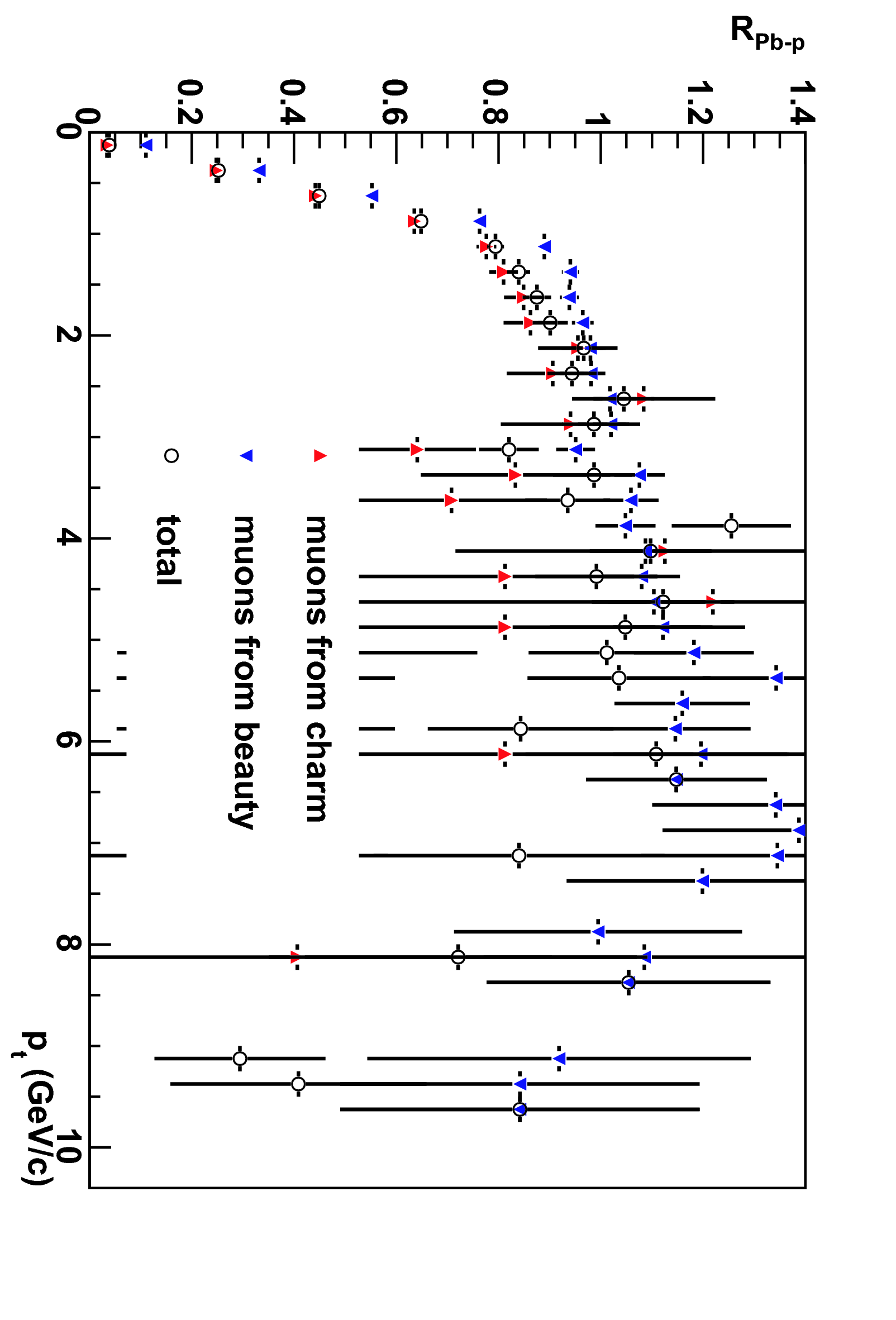}


\caption{$R_{Pbp}$ ratio  for muons from charm and beauty reconstructed in the ALICE muon
spectroometer.
  The left (right) plot corresponds to the results obtained with the CGC (MNR) model.} \label{Fig2}
\end{figure}

The ratio $R_{Pbp}$, presented in Fig.~\ref{Fig2} as a function of $p_{t}$ for muons from charm
and beauty, exhibits a similar behavior for both flavours in the two models. The observation of
the saturation effects via the $R_{Pbp}$ ratio will be difficult at low $p_{t}$
($p_{t}~<~$2~GeV/c) in the ALICE muon spectrometer due to the different experimental cuts.


\section{Conclusions}
The main purpose of this work was to investigate the predictions of the CGC approach in the case
of charm and beauty production rates for the ALICE experiment. The results were compared to the
reference simulations of ALICE which are based on the MNR calculation. It was observed that the
heavy quark production is affected by small-$x$ effects in the low $p_{t}$ region. It was shown
that the shadowing effects are much more significant in the case of the CGC approach. It was also
seen that the Cronin effect is very important in the CGC framework, especially for charm which is
more affected because of its smaller mass.\\
The muon level results were obtained from realistic simulations containing the muon spectrometer
response function. Finally, we conclude that to observe experimentally small-$x$ effects with the
muon spectrometer of the ALICE detector a very challenging analysis has to be done.


\begin{thebibliography}{99}
\bibitem{DGLAP}
 V.N.~Gribov and L.N.~Lipatov, {\it Sov.\ Journ.\ Nucl.\ Phys.}\ {\bf 15}
  (1972), 438; G. Altarelli and G. Parisi, {\it Nucl.\ Phys.}\,{\bf B126}
  (1977), 298; Yu. L.~Dokshitzer, {\it Sov.\ Phys.\ JETP} {\bf 46} (1977), 641.

\bibitem{BFKL}
 L.N.~Lipatov, {\it Sov.\ J.\ Nucl.\ Phys.}\,{\bf 23} (1976) 338;
  E.A.~Kuraev, L.N.~Lipatov and V.S.~Fadin, {\it Zh. Eksp. Teor. Fiz} {\bf 72},
  3 (1977) ({\it Sov. Phys. JETP }{\bf 45} (1977) 199); Ya.Ya.~Balitsky and
  L.N.~Lipatov, {\it Sov.\ J.\ Nucl.\ Phys.} {\bf 28} (1978) 822.

\bibitem{CGCreviews}
E.~Iancu and R.~Venugopalan,  hep-ph/0303204.


\bibitem{ALICE}
ALICE Collaboration, {\it J.\ of\ Physics\ G\ Nucl.\ Part.\ Phys.}\,{\bf 32} (2006) 1295.

\bibitem{aliroot}
{\it http://aliceinfo.cern.ch/Offline/AliRoot/Manual.html}

\bibitem{MNR}
 M.L.~Mangano, P.~Nason and G.~Ridol, {\it Nucl\ Phys.}\,{\bf B}{\bf 373} (1992) 295.

\bibitem{muon}
ALICE Collaboration, {\it JINST}\,{\bf 3}, S08002 (2008); ALICE Collaboration, {\it
CERN/LHCC}\,~{\bf 99-22}, (1999); ALICE Collaboration, {\it CERN/LHCC}\,~{\bf 2000-046}, (2000).

\bibitem{PumplBHST1}
J.~Pumplin, A.~Belyaev, J.~Huston, D.~Stump, W.K.~Tung, \textit{JHEP} {\bf 0602}, (2006) 032.
\bibitem{glau}
R.J.~Glauber and G.~Matthiae, {\it Nucl\ Phys.}\,{\bf B}{\bf 21} (1970) 135.

\bibitem{EskolKS2}
K.J.~Eskola, V.J.~Kolhinen, C.A.~Salgado, \textit{Nucl.\ Phys.} {\bf A} {\bf 661}, (1999) 645.

\bibitem{McLerV1}
L.D.~McLerran, R.~Venugopalan, \textit{Phys.\ Rev.} {\bf D} {\bf 49}, (1994) 2233.

\bibitem{transport}
L.~Aphecetche et al., \textit{ ALICE-Note to be published}.


\end{thebibliography}
\end{document}